\documentclass[useAMS,usenatbib]{mn2e}
\usepackage{graphicx}
\usepackage{txfonts}
\usepackage{natbib}
\usepackage{hyperref}
\usepackage{times,amssymb,amsfonts,amsbsy}

\begin{document}

\title[Angular momentum growth]
  {Revisiting the angular momentum growth of protostructures evolved from non-Gaussian initial conditions}

\author[C.\,Fedeli]{C.\,Fedeli\\
      Department of Astronomy, University of Florida, 211 Bryant Space Science Center, Gainesville, FL 32611 (\sc{cosimo.fedeli@astro.ufl.edu})}
      
\maketitle

\begin{abstract}
I adopt a formalism previously developed by Catelan and Theuns (CT) in order to estimate the impact of primordial non-Gaussianity on the quasi-linear spin growth of cold dark matter protostructures. A variety of bispectrum shapes are considered, spanning the currently most popular early Universe models for the occurrence of non-Gaussian density fluctuations. In their original work, CT considered several other shapes, and suggested that only for one of those does the impact of non-Gaussianity seem to be perturbatively tractable. For that model, and on galactic scales, the next-to-linear non-Gaussian contribution to the angular momentum variance has an upper limit of $\sim 10\%$ with respect to the linear one. I find that all the new models considered in this work can also be seemingly described via perturbation theory. Considering current bounds on $f_\mathrm{NL}$ for inflationary non-Gaussianity leads to the quasi-linear contribution being $\sim 10-20\%$ of the linear one. This result motivates the systematic study of higher-order non-Gaussian corrections, in order to attain a comprehensive picture of how structure gravitational dynamics descends from the physics of the primordial Universe.
\end{abstract}

\section{Introduction}

Gravitational instability and hierarchical growth of Cold Dark Matter (CDM) density perturbations provide an elegant description for the acquisition of angular momentum by protostructures. Accordingly, patches of matter are spun up by tidal torques exerted by the surrounding Large-Scale Structure (LSS, see \citealt{PE69.1,PE71.1,DO70.1,WH84.1,HE88.1}). At the linear level the spin growth is given by the coupling of the first-order deformation tensor and the inertia tensor of the patch, and it agrees quite well with more accurate numerical results (see the review by \citealt{SC09.1} and the references therein).

Acquisition of angular momentum beyond the simple linear description has been tackled by means of Lagrangian perturbation theory in a series of papers by Catelan and Theuns (\citealt*{CA95.1,CA96.1,CA96.2,CA97.1}, CT henceforth) almost two decades ago. In particular, they found that the next-to-linear correction to the growth of ensemble-averaged spin is non-vanishing only if primordial density fluctuations are non-Gaussian. CT explored several non-Gaussian models, and concluded that only for one of these does the angular momentum acquisition appear perturbatively tractable: the log-normal model for the gravitational potential of \citet{MO91.1}. For this template, and adopting a representative mass scale of $M\sim 10^{12} h^{-1}M_\odot$ (with $h=0.5$ and Gaussian filtering), CT found an upper limit of $\sim 24\%$ for the quasi-linear non-Gaussian contribution to the spin variance. This figure translates to a $\sim10\%$ value when rescaled to a smaller and more typical galactic mass of $M=10^{10}h^{-1}M_\odot$ (with $h=0.7$ and a real-space top-hat filter). For other templates the non-Gaussian contribution is comparable to, or larger than, the linear term, suggesting the impossibility of a perturbative expansion.

Since then, the issue of angular momentum growth in non-Gaussian cosmologies has not been investigated further. On the contrary, 
new and more general models of primordial non-Gaussianity exist nowadays and, most importantly, constraints on the level of primordial non-Gaussianity coming from the Cosmic Microwave Background (CMB) and the LSS have dramatically improved over the last decade. Given the cosmological relevance of primordial non-Gaussianity (see \citealt{BA04.1,CH10.1} for recent reviews) and the significance of the CDM halo angular momentum acquisition for the formation and evolution of galaxies, it is important to update this topic. Specifically, it is interesting to explore the amplitudes and behaviors of the quasi-linear contributions to spin growth given by non-Gaussian models that are popular nowadays. This is the scope of the present letter.

The rest of the manuscript is organized as follows. In Section \ref{sct:acquisition} I review the linear and next-to-linear contributions to the ensemble-averaged spin growth of matter patches. In Section \ref{sct:nongaussian} I summarize the non-Gaussian cosmologies explored here. In Section \ref{sct:results} results are displayed and in Section \ref{sct:conclusions} conclusions are drawn. Where needed, I adopted the following cosmological parameters: $\Omega_{\mathrm{m},0} = 0.272$, $\Omega_{\Lambda,0} = 1-\Omega_{\mathrm{m},0}$, $\Omega_{\mathrm{b},0} = 0.046$, $H_0 = 100 h $ km s$^{-1}$ Mpc$^{-1}$ with $h = 0.704$, $\sigma_8 = 0.809$, and $n_\mathrm{s}=1$.

\section{Angular momentum acquisition}\label{sct:acquisition}

\subsection{Lagrangian displacement}

In Lagrangian theory the comoving position $\boldsymbol x$ of a mass element at time $\tau$ can be written in terms of its initial position $\boldsymbol q$ and a displacement vector field $\boldsymbol S$, as

\begin{equation}
\boldsymbol x (\boldsymbol q, \tau) = \boldsymbol q + \boldsymbol S(\boldsymbol q, \tau)~.
\end{equation}
Following CT here I used a time variable $\tau$ that is related to the standard cosmic time $t$ by $d\tau = dt/a^2$ \citep{SH80.1}, where $a$ is the scale factor.

Perturbative approximations to this exact expression can be found by expanding the displacement field in a series,

\begin{equation}\label{eqn:expansion}
\boldsymbol S = \sum_{n=1}^\infty \boldsymbol S_n~,
\end{equation}
where $\boldsymbol S_1$ corresponds to the Zel'dovich approximation $\boldsymbol S_1(\boldsymbol q, \tau) = D(\tau) \nabla \psi_1(\boldsymbol q)$. Here $D(\tau)$ is the growth factor of linear density perturbations, which in a Einstein-de Sitter universe reads $D(\tau) = \tau^{-2}$. The function $\psi_1$ is the first order (Zel'dovich) displacement potential \citep{ZE70.1}, related to the linear density perturbation field by the Poisson equation $\Delta \psi_1(\boldsymbol q) = \delta(\boldsymbol q)$, so that in Fourier space $\hat\psi_1(\boldsymbol p) = \hat\delta(\boldsymbol p)/p^2$.

The second-order term of the displacement field can also be separated in time and space, according to $\boldsymbol S_2(\boldsymbol q, \tau) = E(\tau)\nabla\psi_2(\boldsymbol q)$. The growth factor $E(\tau)$ reads $E(\tau) = -3\tau^{-4}/7$ in an Einstein-de Sitter universe, while for its more general expression I refer to CT. The second-order displacement potential can be related to its first-order counterpart in Fourier space by

\begin{eqnarray}
\hat\psi_2(\boldsymbol p)&=&-\frac{1}{p^2}\int_{\mathbb{R}^6} \frac{\mathrm{d}\boldsymbol p_1\mathrm{d}\boldsymbol p_2}{(2\pi)^6} \left[(2\pi)^3\delta_\mathrm{D}(\boldsymbol p_1+\boldsymbol p_2 - \boldsymbol p) \right] K(\boldsymbol p_1,\boldsymbol p_2)\times
\nonumber\\
&\times& \hat\psi_1(\boldsymbol p_1)\hat\psi_1(\boldsymbol p_2)~.
\end{eqnarray}
In the previous Equation $K(\boldsymbol p_1,\boldsymbol p_2)$ is a symmetric integration kernel defined as

\begin{equation}
K(\boldsymbol p_1,\boldsymbol p_2) \equiv \frac{1}{2} \left[ p_1^2p_2^2 - \left(\boldsymbol p_1\cdot \boldsymbol p_2\right)^2 \right] = \frac{1}{2}p_1^2p_2^2\left(1-\mu^2\right)~,
\end{equation}
where $\mu$ is the cosine of the angle between the two wavevectors $\boldsymbol p_1$ and $\boldsymbol p_2$.

\subsection{Spin growth}

The angular momentum of the matter initially contained in a comoving Lagrangian patch $\Gamma$ of the Universe at time $\tau$ can be written as an integral over $\Gamma$,

\begin{equation}
\boldsymbol J(\tau) = a^3(\tau) \rho_{\mathrm{m},0}\int_{\Gamma} \mathrm{d}\boldsymbol q \left[ \boldsymbol q + \boldsymbol S(\boldsymbol q,\tau) \right] \times \frac{\partial\boldsymbol S(\boldsymbol q, \tau)}{\partial\tau}~.
\end{equation}
By considering the series expansion of the Lagrangian displacement field $\boldsymbol S$ introduced in Eq. (\ref{eqn:expansion}), the angular momentum of the patch can be similarly written as 

\begin{equation}
\boldsymbol J = \sum_{m=1}^\infty \boldsymbol J_m~.
\end{equation}
The first-order term of the angular momentum series takes the form

\begin{equation}
\boldsymbol J_1(\tau) = a^3(\tau) \rho_{\mathrm{m},0} \frac{dD(\tau)}{d\tau} \int_{\Gamma} \mathrm{d}\boldsymbol q~\boldsymbol q \times \nabla\psi_1(\boldsymbol q)~.
\end{equation}

By expanding the Zel'dovich potential around the center of mass of the patch (assumed to be, without loss of generality, the origin of the reference frame) up to the second order, the previous equation takes the compact form

\begin{equation}
J_{1,\alpha}(\tau) = \frac{dD(\tau)}{d\tau}\varepsilon_{\alpha\beta\gamma}\mathcal{D}_{1,\beta\sigma}\mathcal{I}_{\sigma\gamma}(\tau)~.
\end{equation}
In the previous equation $\varepsilon_{\alpha\beta\gamma}$ is the fully antisymmetric Levi-Civita tensor, $\mathcal{D}_{1,\beta\sigma}$ is the Zel'dovich deformation tensor,

\begin{equation}
\mathcal{D}_{1,\beta\sigma} \equiv \frac{\partial^2\psi_1(\boldsymbol 0)}{\partial q_\beta\partial q_\sigma} =  -\int_{\mathbb{R}^3}\frac{\mathrm{d}\boldsymbol p}{(2\pi)^3}~p_\beta p_\sigma \hat\psi_1(\boldsymbol p)~,
\end{equation}
while $\mathcal{I}_{\sigma\gamma}$ is the inertia tensor of the patch,

\begin{equation}
\mathcal{I}_{\sigma\gamma}(\tau) \equiv a^3(\tau)\rho_{\mathrm{m},0} \int_{\Gamma} \mathrm{d}\boldsymbol q~q_\sigma q_\gamma~.
\end{equation}
Summation over repeated indices is implicit.

Likewise, the second-order term in the series expansion of the angular momentum reads

\begin{equation}
\boldsymbol J_2(\tau) = a^3(\tau) \rho_{\mathrm{m},0} \frac{dE(\tau)}{d\tau} \int_{\Gamma} \mathrm{d}\boldsymbol q~\boldsymbol q \times \nabla\psi_2(\boldsymbol q)~,
\end{equation}
which, under a second-order Taylor expansion of the displacement potential takes the form 

\begin{equation}
J_{2,\alpha}(\tau) = \frac{dE(\tau)}{d\tau}\varepsilon_{\alpha\beta\gamma}\mathcal{D}_{2,\beta\sigma}\mathcal{I}_{\sigma\gamma}(\tau)~.
\end{equation}
It can be shown that the second-order deformation tensor in Fourier space reads

\begin{eqnarray}
\mathcal{D}_{2,\beta\sigma} &=& \int_{\mathbb{R}^6}\frac{\mathrm{d}\boldsymbol p_1\mathrm{d}\boldsymbol p_2}{(2\pi)^6} \frac{(\boldsymbol p_1+\boldsymbol p_2)_\beta(\boldsymbol p_1+\boldsymbol p_2)_\sigma}{\|\boldsymbol p_1+\boldsymbol p_2\|^2} K(\boldsymbol p_1,\boldsymbol p_2)\times
\nonumber\\
&\times& \hat\psi_1(\boldsymbol p_1) \hat\psi_1(\boldsymbol p_2)~,
\end{eqnarray}
in terms of the Zel'dovich potential.

\subsection{Ensemble averages}

In order to simplify the previous results, it is meaningful to consider the ensemble average of the square of the angular momentum. It then follows that, up to the next-to-linear order,

\begin{equation}
\left \langle \|\boldsymbol J(\tau)\|^2 \right \rangle \simeq \left \langle \|\boldsymbol J_1(\tau)\|^2 \right \rangle + 2\left \langle \boldsymbol J_1(\tau)\cdot \boldsymbol J_2(\tau) \right \rangle~,
\end{equation}
where

\begin{equation}\label{eqn:gaussian}
\left \langle \|\boldsymbol J_1(\tau)\|^2 \right \rangle = \frac{2}{15}\left[\frac{dD(\tau)}{d\tau}\right]^2\left( \nu_1^2-3\nu_2 \right)\sigma_M^2~.
\end{equation}
In the previous Equation $\sigma_M$ is the mean deviation of the matter density field smoothed on a scale corresponding to mass $M$, while $\nu_1$ and $\nu_2$ are the first and second invariant of the inertia tensor, respectively. To be more precise, if $\lambda_1$, $\lambda_2$, and $\lambda_3$ are the three eigenvalues of the inertia tensor, then $\nu_1\equiv \lambda_1+\lambda_2+\lambda_3$ and $\nu_2 \equiv \lambda_1\lambda_2+\lambda_1\lambda_3 + \lambda_2\lambda_3$.

The next term is non-vanishing only if density fluctuations are non-Gaussian, and reads

\begin{equation}\label{eqn:nongaussian}
\left \langle \boldsymbol J_1(\tau)\cdot \boldsymbol J_2(\tau) \right \rangle = \frac{2}{15} \frac{dD(\tau)}{d\tau}\frac{dE(\tau)}{d\tau} \left( \nu_1^2-3\nu_2\right) \omega_M~,
\end{equation}
where 

\begin{eqnarray}\label{eqn:omega}
\omega_M &=& -15\int_{\mathbb{R}^6} \frac{\mathrm{d}\boldsymbol p_1\mathrm{d}\boldsymbol p_2}{(2\pi)^6}\|\boldsymbol p_1+\boldsymbol p_2\|^2 K(\boldsymbol p_1,\boldsymbol p_2) \hat W^2_R(\|\boldsymbol p_1 + \boldsymbol p_2\|)
\times
\nonumber\\
&\times& B_{\psi_1} (\boldsymbol p_1,\boldsymbol p_2,-\boldsymbol p_1-\boldsymbol p_2)~.
\end{eqnarray}
In the previous Equation $B_{\psi_1}$ represents the bispectrum of the Zel'dovich potential, which is now explicitly smoothed on a scale $R = (2GM/\Omega_{\mathrm{m},0}H_0^2)^{1/3}$. I assumed the standard real-space top-hat smoothing. The Zel'dovich potential can be related to the standard gravitational potential $\varphi$ by making use of the Poisson equation,

\begin{equation}
\hat\psi_1(\boldsymbol p) = \frac{2}{3}\frac{T(p)}{H_0^2\Omega_{\mathrm{m},0}}\hat\varphi (\boldsymbol p)\equiv F(p)\hat\varphi(\boldsymbol p)~,
\end{equation}
where $T(p)$ is the cold dark matter transfer function \citep{BA86.1,SU95.1}. The integral in Eq. (\ref{eqn:omega}) has to be solved numerically for realistic bispectrum shapes. Fortunately, the bispectrum usually depends only on the magnitude of its three arguments, so that the above six-dimensional integral reduces to a three-dimensional one.

\section{Non-Gaussian shapes}\label{sct:nongaussian}

I considered five different shapes for the primordial bispectrum, that are briefly described below. The first four are motivated by inflationary physics, and the amplitude of non-Gaussianity is given by the parameter $f_\mathrm{NL}$ (assumed to be constant). The fifth is non-inflationary in nature, and hence independent on $f_\mathrm{NL}$. See \citet{FE11.1} and references therein.

\subsection{Local shape}

This bispectrum shape arises when a light scalar field, additional to the inflaton, contributes to the curvature perturbations \citep*{BE02.1,BA04.2,SA06.1}. It is the same shape produced by the standard model of inflation \citep*{FA93.1} but in this case the amplitude can be arbitrary. The potential bispectrum takes the simple form

\begin{equation}
B_\varphi(\boldsymbol p_1,\boldsymbol p_2, \boldsymbol p_3) = 2A^2f_\mathrm{NL}\left[ \left(p_1p_2\right)^{n_\mathrm{s}-4} + \left(p_1p_3\right)^{n_\mathrm{s}-4} + \left(p_2p_3\right)^{n_\mathrm{s}-4} \right]~,
\end{equation}
and it is maximized for squeezed configurations. The quantity $A$ is the spectral amplitude of the potential (given by $\sigma_8$), while $n_\mathrm{s}$ is the spectral slope.

\subsection{Equilateral shape}

This shape is a consequence of the inflaton Lagrangian being non-standard, and containing higher-order derivatives of the field (\citealt*{AL04.1}; \citealt{AR04.1}; \citealt*{LI08.1}). The resulting bispectrum is maximized for equilateral configurations. A template for the equilateral bispectrum can be found in \citet{CR07.1}, however the expression is rather cumbersome and I did not report it here. The same applies to the following shapes.

\subsection{Enfolded shape}

The enfolded shape results from primordial non-Gaussianity being evaluated without the regular Bunch-Davies vacuum hypothesis \citep{CH07.1,HO08.1}. In this case the bispectrum is maximized for squashed configurations. A template for such a bispectrum is reported in \citet*{ME09.1}.

\subsection{Orthogonal shape}

This shape is defined as being orthogonal (with respect to a suitably defined scalar product) to both the local and equilateral forms. The resulting bispectrum is maximized for both equilateral and squashed configurations, and a template can be found in \citet*{SE10.1}.

\subsection{Matter bounce shape}

This configuration is the consequence of a model universe without inflation, but with a scale factor that bounces in a non-singular way \citep{BR09.1,CA09.1}. The matter bounce leads to a scale-invariant spectrum of density fluctuations, and to a bispectrum whose shape is similar to the local shape. Being non-inflationary in origin, the non-Gaussianity induced by a matter bounce model has no dependence on $f_\mathrm{NL}$. It can instead be shown that the matter bounce bispectrum is comparable to the local one with a fixed $f_\mathrm{NL} = -35/8$. I considered explicitly the matter bounce because it is in principle possible that a weighted integral of the bispectrum, such as the one in Eq. (\ref{eqn:omega}), would magnify its differences with respect to the local model. As I show below this is actually not the case.

\section{Results}\label{sct:results}

\begin{figure}
\centering
\includegraphics[width=\hsize]{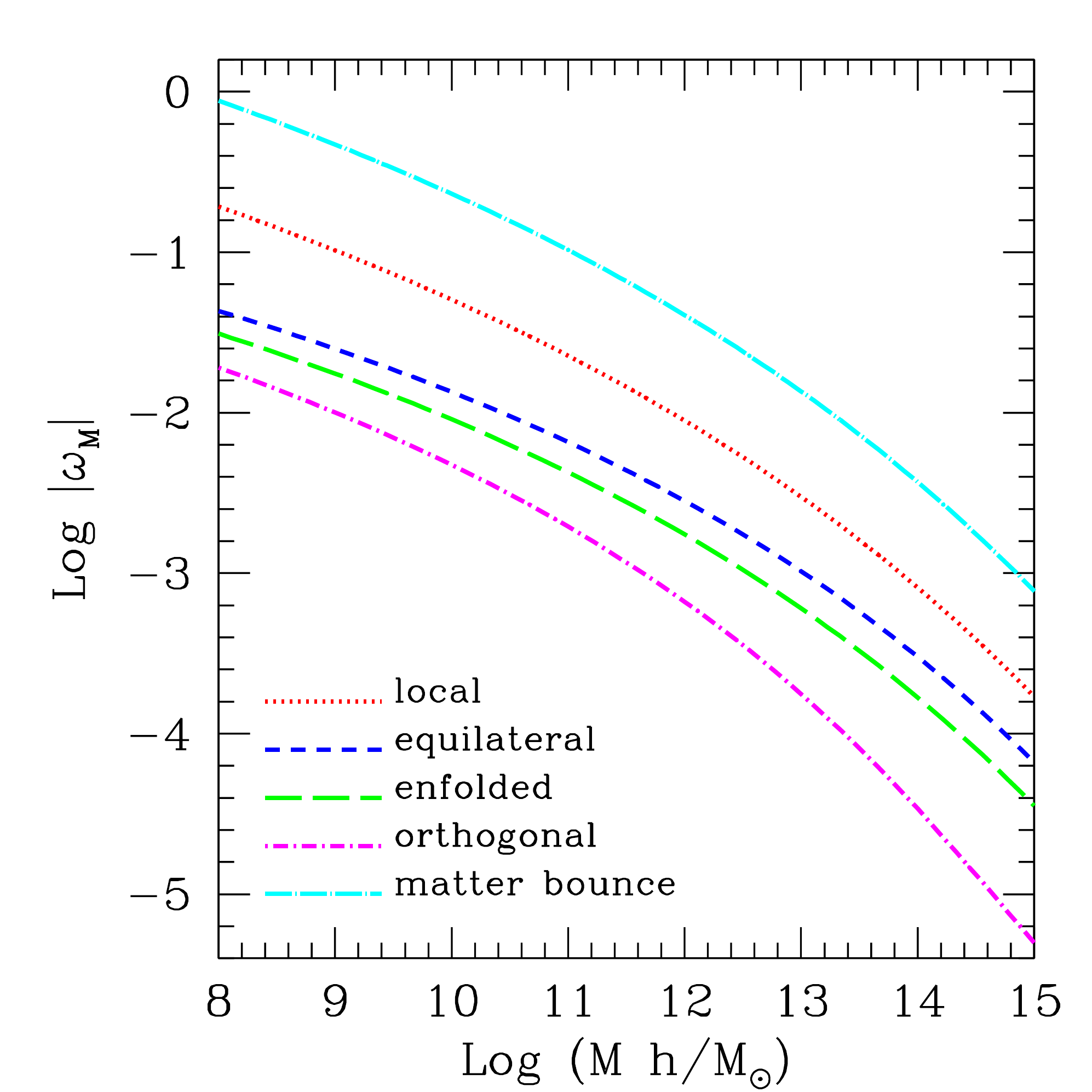}
\caption{The shape of the function $\omega_M$, quantifying the non-Gaussian contribution to the angular momentum variance growth. Different line styles and colors refer to different non-Gaussian bispectrum shapes, as labeled. The non-Gaussianity induced by the matter bounce is independent on $f_\mathrm{NL}$, while in all other cases $f_\mathrm{NL} = 1$ has been assumed.}
\label{fig:omega}
\end{figure}

In Figure \ref{fig:omega} I show $\omega_M$ as a function of the mass scale for the five non-Gaussian cosmologies considered in this letter. In all cases, except for the matter bounce, I selected $f_\mathrm{NL} = 1$ in order to purely highlight the effect of the bispectrum shape. It is however easy to see that $\omega_M$ is simply proportional to $f_\mathrm{NL}$. All curves decrease with increasing mass, implying that larger matter patches acquire lower amounts of angular momentum than smaller ones. This behavior is similar to the mass dependence of the linear term given in Eq. (\ref{eqn:gaussian}). Also, curves referring to different models are rather similar in shape. Besides the matter bounce model, the local model is the one having the largest effect, while the orthogonal model has the lowest. This is different from the behavior of, e.g., the halo bias, for which the equilateral model displays the smallest effect. The function $\omega_M$ for the matter bounce is virtually identical to that for the local model when assuming $f_\mathrm{NL} = -35/8$, in agreement with the previous discussion.

As can be seen from the structure of Eqs. (\ref{eqn:gaussian}) and (\ref{eqn:nongaussian}), the relative importance of the non-Gaussian contribution with respect to the linear one is

\begin{equation}
\Upsilon_M(\tau)\equiv2\frac{\left\langle \boldsymbol J_1(\tau)\cdot \boldsymbol J_2(\tau) \right\rangle}{\left\langle \|\boldsymbol J_1(\tau)\|^2 \right\rangle} = 2\frac{dE(\tau)/d\tau}{dD(\tau)/d\tau} \frac{\omega_M}{\sigma_M^2}~.
\end{equation}
The acquisition of angular momentum by protostructures happens at high redshift where, under the assumption of flat spatial geometry, the Universe is well approximated by an Einstein-de Sitter model. Corrections due to the presence of a cosmological constant can be considered to be negligible. If this is the case, then

\begin{figure}
\centering
\includegraphics[width=\hsize]{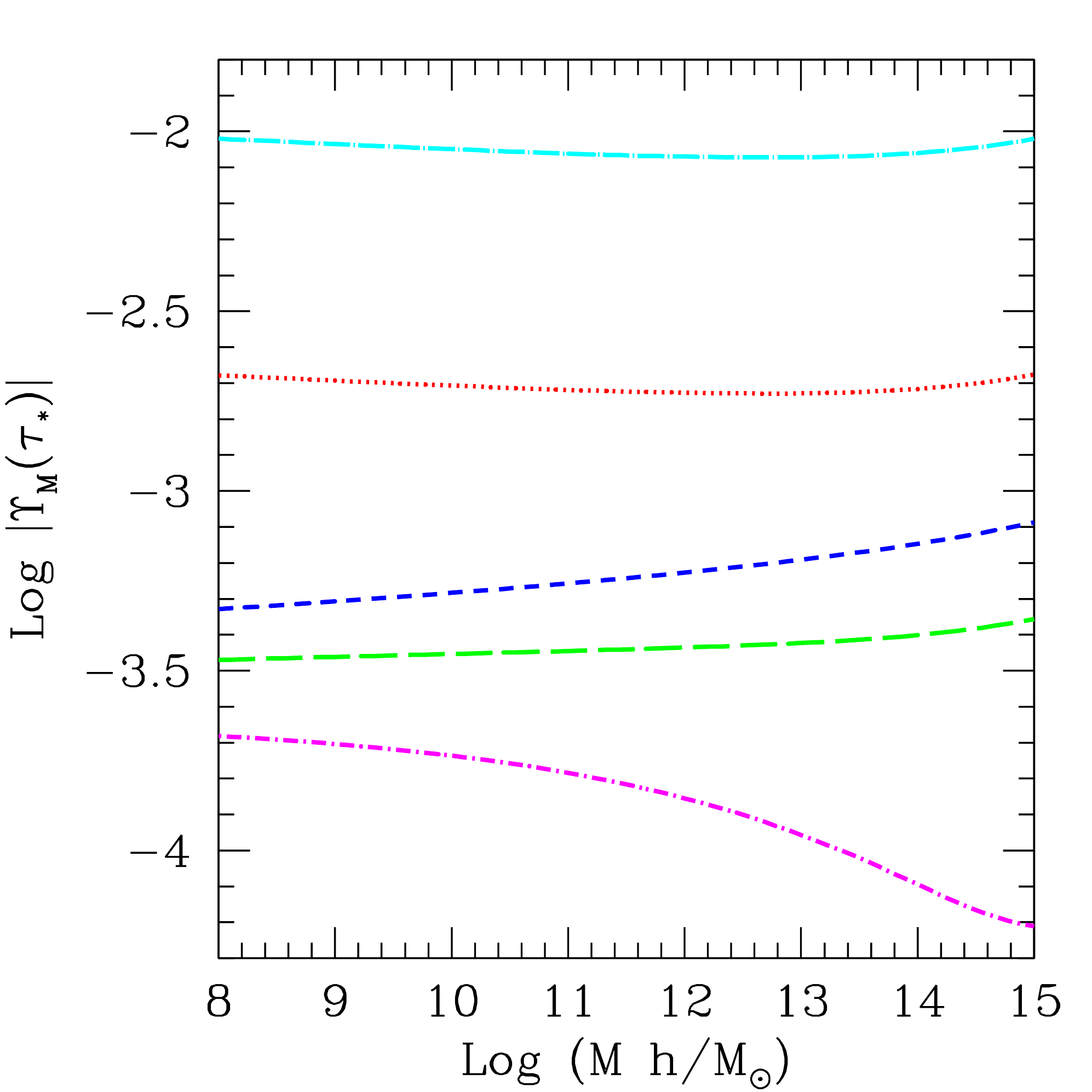}
\caption{The non-Gaussian contribution to the spin variance growth normalized by the linear contribution, evaluated at the collapse time for an overdensity of a given mass. Line types and colors are the same as in Figure \ref{fig:omega}, and $f_\mathrm{NL} = 1$ has been assumed for all models except the matter bounce (which is $f_\mathrm{NL}$-independent).}
\label{fig:upsilon_mass}
\end{figure}

\begin{equation}
\Upsilon_M(\tau) = -\frac{12}{7}\tau^{-2} \frac{\omega_M}{\sigma_M^2} = -\frac{12}{7}(3t)^{2/3}\frac{\omega_M}{\sigma_M^2}~.
\end{equation}
Note that the negative sign cancels with the negative sign in the definition of $\omega_M$, so that the non-Gaussian contribution is positive (increases the spin growth) for a model with positive skewness and negative otherwise. This has already been noticed by CT. The previous equation also shows that the non-Gaussian contribution to spin acquisition grows faster than the linear one.

If the matter patch under consideration is an overdense region, it is reasonable to assume that the spin growth induced by tidal torques occurs until the overdensity detaches from the overall expansion of the Universe and collapses into a bound structure. This moment $\tau_*$ can be naively identified as $D(\tau_*)\sigma_M = 1$. For an Einstein-de Sitter cosmological model this implies $\tau_*^2 = \sigma_M$, and thus

\begin{equation}
\Upsilon_M(\tau_*) = -\frac{12}{7} \frac{\omega_M}{\sigma_M^3}~.
\end{equation}

In Figure \ref{fig:upsilon_mass} I show the mass dependence of the function $\Upsilon_M(\tau_*)$ for the various non-Gaussian cosmologies that have been considered in this work. For models with inflationary non-Gaussianity I assumed $f_\mathrm{NL} = 1$. As can be seen the mass dependence is in all cases relatively weak. In the local and matter bounce models $\Upsilon_M(\tau_*)$ is basically unchanged for masses ranging between $M = 10^8 h^{-1} M_\odot$ and $M = 10^{15} h^{-1} M_\odot$. For the equilateral and enfolded models $\Upsilon_M(\tau_*)$ increases by $\sim 50\%$ over the same interval, while for the orthogonal case it decreases by a factor of $\sim 3$. Hence, despite the fact that the linear and non-Gaussian contributions to the spin growth both decrease with mass, their relative importance remains relatively unchanged. The only exception is represented by the orthogonal model.

\begin{figure}
\centering
\includegraphics[width=\hsize]{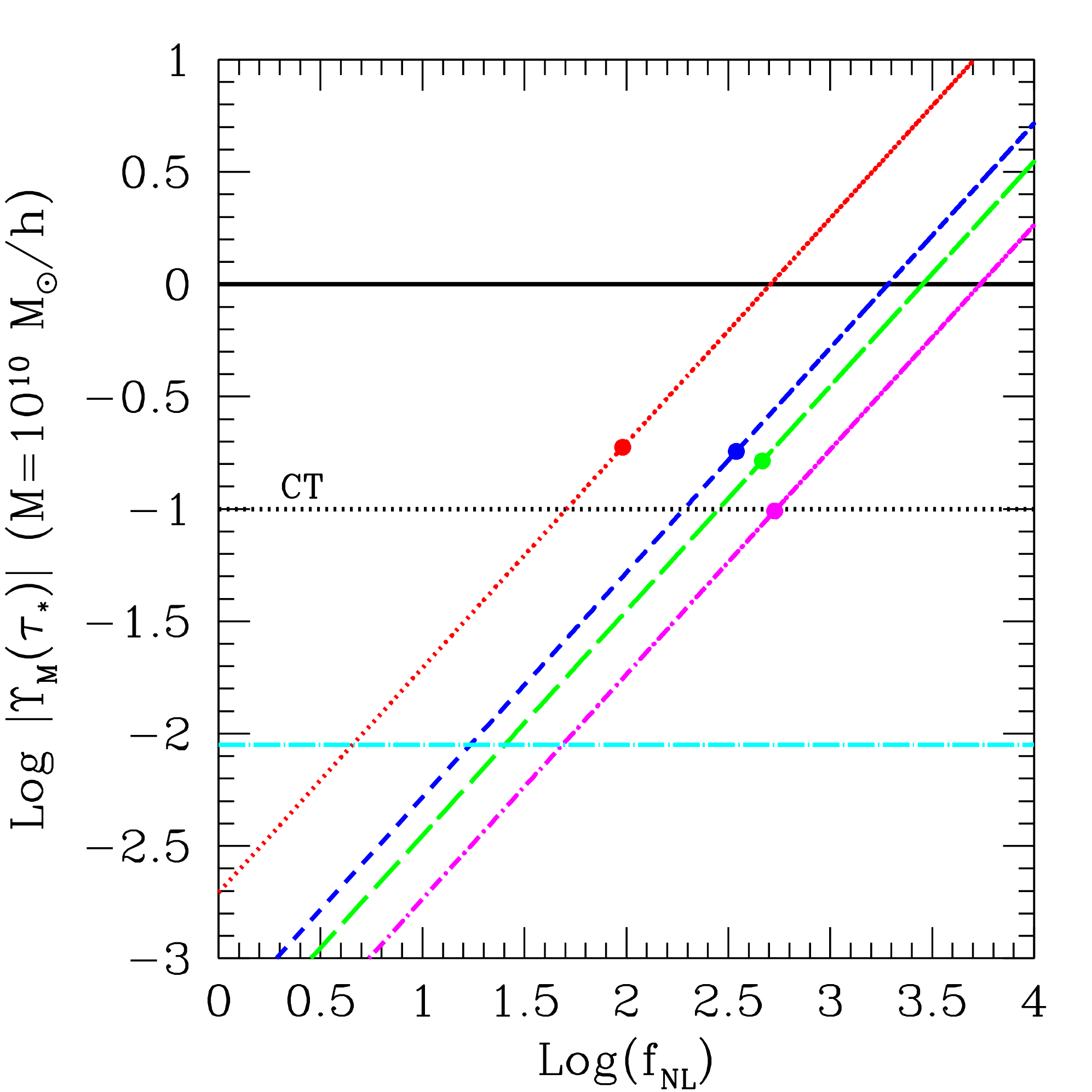}
\caption{The non-Gaussian contribution to the spin variance growth normalized by the linear contribution, evaluated at the collapse time for an overdensity of mass $M=10^{10} h^{-1} M_\odot$. Line types and colors are the same as in Figure \ref{fig:omega}, and results are shown as a function of $f_\mathrm{NL}$. The black solid line highlights the locus where the non-Gaussian contribution is identical to the linear one. Filled circles on each curve represent the maximum $|f_\mathrm{NL}|$ values allowed by current constraints from CMB and LSS. The black dotted line shows the upper limit to the non-Gaussian spin contribution found by CT assuming a log-normal model for the primordial gravitational potential, after rescaling CT's result as detailed in the text.}
\label{fig:upsilon_fnl}
\end{figure}

Next, I selected a reference mass scale of $M = 10^{10} h^{-1} M_\odot$ and computed the dependence of $\Upsilon_M(\tau_*)$ on $f_\mathrm{NL}$, shown in Figure \ref{fig:upsilon_fnl}. As previously mentioned, this dependence is always linear, however the Figure is important in order to understand for what value of $f_\mathrm{NL}$ a certain non-Gaussian model provides a given contribution to the total angular momentum variance. The matter bounce non-Gaussianity is independent of $f_\mathrm{NL}$, hence its contribution is always at the percent (negative) level compared to the linear one. As for the other models, in order for the non-Gaussian contribution to be comparable to the linear one, primordial non-Gaussianity would need to be at the unrealistic level of $f_\mathrm{NL}\sim 400$ for the local case, and substantially larger than that for other bispectrum shapes.

Figure \ref{fig:upsilon_fnl} allows one to determine the non-Gaussian contribution, in units of the linear contribution, for the current bounds on $f_\mathrm{NL}$. Constraints from the CMB \citep{KO11.1} imply $-13 < f_\mathrm{NL} < 96$ at $95\%$ Confidence Level (CL) for the local shape\footnote{I converted all constraints mentioned here to the LSS convention, which has been adopted throughout. See \citet{FE10.1} and references therein for a discussion.}, meaning that the non-Gaussian contribution can be at most $\sim 19\%$ of the linear one. The same CMB data constrain $-278 < f_\mathrm{NL} < 346$ for the equilateral shape, implying a $\sim 18\%$ relative importance. The tighter constraints on the level of non-Gaussianity for the enfolded shape come from the LSS \citep{XI11.1}, corresponding to $-16 < f_\mathrm{NL} < 465$ at $2\sigma$ confidence level. This means that the non-Gaussian contribution is at most $\sim 16\%$ of the linear one. Finally, for the orthogonal shape the CMB data by \citet{KO11.1} bear $-533 < f_\mathrm{NL} < 8$, implying a maximum $\sim 10\%$ (negative) relative strength. These numbers can be appreciated also by looking at the positions of the filled circles in Figure \ref{fig:upsilon_fnl}. For comparison, the black dotted line shows the upper limit to the non-Gaussian contribution found by CT, after assuming a log-normal distribution for the primordial gravitational potential and after rescaling it to the scale $M = 10^{10}h^{-1}M_\odot$. I stress the fact that, while CT adopted a Gaussian window function and $h=0.5$, I used a real-space top-hat filter and $h=0.704$. Moreover, while CT calibrated the level of primordial non-Gaussianity using a value $S_R = 4$ for the skewness of the matter density field on a scale $R=8h^{-1}$Mpc, I adopted, conservatively, $S_R=0.1$. This value results from the large-scale skewness per unit $f_\mathrm{NL}$ for local non-Gaussianity ($\sim 10^{-3}$, e.g., Figure 1 of \citealt{FE11.1}) multiplied by the most recent upper limit on the level of non-Gaussianity for the same shape ($f_\mathrm{NL}\sim 100$).

\section{Conclusions}\label{sct:conclusions}

I reconsidered the impact of primordial non-Gaussianity on the acquisition of angular momentum by CDM protostructures. Non-Gaussian initial conditions provide a next-to-linear correction to the spin growth that is absent when density fluctuations are normally distributed. Previous results, obtained by CT after assuming a log-normal primordial gravitational potential, resulted in a contribution to the spin variance of $\sim 10\%$ with respect to the linear one. This value holds for a scale $M=10^{10}h^{-1}M_\odot$ (with the cosmology of this work) and is based on a matter skewness of $S_R\sim 0.1$, as deduced in the previous Section. Other models turned out to give a very large quasi-linear effect, suggesting that Lagrangian perturbation theory could not be successfully applied in those cases. I found that for several current models of non-Gaussian initial conditions, the contribution to the galactic spin variance during the mildly non-linear regime is similar to what predicted assuming a log-normal primordial gravitational potential. Considering the upper limits to the current constraints on the level of inflationary primordial non-Gaussianity returns a next-to-linear contribution at the level of $\sim 10-20\%$. These results imply that the spin growth induced by inflationary non-Gaussianity seems to be generically tractable via perturbation theory.

CT also demonstrated that higher-order contributions in the case of Gaussian density fluctuations provide a correction to the angular momentum variance equal to $\sim 60\%$ of the linear term. This means that the next-to-linear non-Gaussian contribution estimated here has a significant impact on the spin acquisition by protostructures. Such an impact could potentially be even larger, because higher-order non-Gaussian corrections, that have not been considered here, depend on the trispectrum of the Zel'dovich potential, and hence also react to primordial non-Gaussianity. The results presented in this letter motivate the study of these higher-order contributions, and show how it is possible to consistently describe the dynamics of protostructures based on largely general cosmological initial conditions.
\vspace{-0.075cm}
\section*{Acknowledgements}

I thank the University of Florida for support through the Theoretical Astrophysics Fellowship. I credit L. Moscardini for insightful comments on the manuscript and I am deeply indebted to the anonymous referee for help in substantially improving this work.
\vspace{-0.075cm}
\small
\bibliographystyle{aa}
\bibliography{master}

\begin{thebibliography}{35}
\expandafter\ifx\csname natexlab\endcsname\relax\def\natexlab#1{#1}\fi

\bibitem[{{Alishahiha} {et~al.}(2004){Alishahiha}, {Silverstein}, \&
  {Tong}}]{AL04.1}
{Alishahiha}, M., {Silverstein}, E., \& {Tong}, D. 2004, \prd, 70, 123505

\bibitem[{{Arkani-Hamed} {et~al.}(2004){Arkani-Hamed}, {Creminelli},
  {Mukohyama}, \& {Zaldarriaga}}]{AR04.1}
{Arkani-Hamed}, N., {Creminelli}, P., {Mukohyama}, S., \& {Zaldarriaga}, M.
  2004, Journal of Cosmology and Astro-Particle Physics, 4, 1

\bibitem[{{Babich} {et~al.}(2004){Babich}, {Creminelli}, \&
  {Zaldarriaga}}]{BA04.2}
{Babich}, D., {Creminelli}, P., \& {Zaldarriaga}, M. 2004, Journal of Cosmology
  and Astro-Particle Physics, 8, 9

\bibitem[{{Bardeen} {et~al.}(1986){Bardeen}, {Bond}, {Kaiser}, \&
  {Szalay}}]{BA86.1}
{Bardeen}, J.~M., {Bond}, J.~R., {Kaiser}, N., \& {Szalay}, A.~S. 1986, \apj,
  304, 15

\bibitem[{{Bartolo} {et~al.}(2004){Bartolo}, {Komatsu}, {Matarrese}, \&
  {Riotto}}]{BA04.1}
{Bartolo}, N., {Komatsu}, E., {Matarrese}, S., \& {Riotto}, A. 2004, \physrep,
  402, 103

\bibitem[{{Bernardeau} \& {Uzan}(2002)}]{BE02.1}
{Bernardeau}, F. \& {Uzan}, J. 2002, \prd, 66, 103506

\bibitem[{{Brandenberger}(2009)}]{BR09.1}
{Brandenberger}, R. 2009, \prd, 80, 043516

\bibitem[{{Cai} {et~al.}(2009){Cai}, {Xue}, {Brandenberger}, \&
  {Zhang}}]{CA09.1}
{Cai}, Y.-F., {Xue}, W., {Brandenberger}, R., \& {Zhang}, X. 2009, \jcap, 5, 11

\bibitem[{{Catelan}(1995)}]{CA95.1}
{Catelan}, P. 1995, \mnras, 276, 115

\bibitem[{{Catelan} \& {Theuns}(1996{\natexlab{a}})}]{CA96.1}
{Catelan}, P. \& {Theuns}, T. 1996{\natexlab{a}}, \mnras, 282, 436

\bibitem[{{Catelan} \& {Theuns}(1996{\natexlab{b}})}]{CA96.2}
{Catelan}, P. \& {Theuns}, T. 1996{\natexlab{b}}, \mnras, 282, 455

\bibitem[{{Catelan} \& {Theuns}(1997)}]{CA97.1}
{Catelan}, P. \& {Theuns}, T. 1997, \mnras, 292, 225

\bibitem[{{Chen}(2010)}]{CH10.1}
{Chen}, X. 2010, Advances in Astronomy, 2010

\bibitem[{{Chen} {et~al.}(2007){Chen}, {Huang}, {Kachru}, \& {Shiu}}]{CH07.1}
{Chen}, X., {Huang}, M., {Kachru}, S., \& {Shiu}, G. 2007, Journal of Cosmology
  and Astro-Particle Physics, 1, 2

\bibitem[{{Creminelli} {et~al.}(2007){Creminelli}, {Senatore}, {Zaldarriaga},
  \& {Tegmark}}]{CR07.1}
{Creminelli}, P., {Senatore}, L., {Zaldarriaga}, M., \& {Tegmark}, M. 2007,
  Journal of Cosmology and Astro-Particle Physics, 3, 5

\bibitem[{{Doroshkevich}(1970)}]{DO70.1}
{Doroshkevich}, A.~G. 1970, Astrophysics, 6, 320

\bibitem[{{Falk} {et~al.}(1993){Falk}, {Rangarajan}, \& {Srednicki}}]{FA93.1}
{Falk}, T., {Rangarajan}, R., \& {Srednicki}, M. 1993, \apjl, 403, L1

\bibitem[{{Fedeli} {et~al.}(2011){Fedeli}, {Carbone}, {Moscardini}, \&
  {Cimatti}}]{FE11.1}
{Fedeli}, C., {Carbone}, C., {Moscardini}, L., \& {Cimatti}, A. 2011, \mnras,
  414, 1545

\bibitem[{{Fedeli} \& {Moscardini}(2010)}]{FE10.1}
{Fedeli}, C. \& {Moscardini}, L. 2010, \mnras, 405, 681

\bibitem[{{Heavens} \& {Peacock}(1988)}]{HE88.1}
{Heavens}, A. \& {Peacock}, J. 1988, \mnras, 232, 339

\bibitem[{{Holman} \& {Tolley}(2008)}]{HO08.1}
{Holman}, R. \& {Tolley}, A.~J. 2008, Journal of Cosmology and Astro-Particle
  Physics, 5, 1

\bibitem[{{Komatsu} {et~al.}(2011){Komatsu}, {Smith}, {Dunkley}, {Bennett},
  {Gold}, {Hinshaw}, {Jarosik}, {Larson}, {Nolta}, {Page}, {Spergel},
  {Halpern}, {Hill}, {Kogut}, {Limon}, {Meyer}, {Odegard}, {Tucker}, {Weiland},
  {Wollack}, \& {Wright}}]{KO11.1}
{Komatsu}, E., {Smith}, K.~M., {Dunkley}, J., {et~al.} 2011, \apjs, 192, 18

\bibitem[{{Li} {et~al.}(2008){Li}, {Wang}, \& {Wang}}]{LI08.1}
{Li}, M., {Wang}, T., \& {Wang}, Y. 2008, Journal of Cosmology and
  Astro-Particle Physics, 3, 28

\bibitem[{{Meerburg} {et~al.}(2009){Meerburg}, {van der Schaar}, \&
  {Corasaniti}}]{ME09.1}
{Meerburg}, P.~D., {van der Schaar}, J.~P., \& {Corasaniti}, S.~P. 2009,
  Journal of Cosmology and Astro-Particle Physics, 5, 18

\bibitem[{{Moscardini} {et~al.}(1991){Moscardini}, {Matarrese}, {Lucchin}, \&
  {Messina}}]{MO91.1}
{Moscardini}, L., {Matarrese}, S., {Lucchin}, F., \& {Messina}, A. 1991,
  \mnras, 248, 424

\bibitem[{{Peebles}(1969)}]{PE69.1}
{Peebles}, P.~J.~E. 1969, \apj, 155, 393

\bibitem[{{Peebles}(1971)}]{PE71.1}
{Peebles}, P.~J.~E. 1971, \aap, 11, 377

\bibitem[{{Sasaki} {et~al.}(2006){Sasaki}, {V{\"a}liviita}, \&
  {Wands}}]{SA06.1}
{Sasaki}, M., {V{\"a}liviita}, J., \& {Wands}, D. 2006, \prd, 74, 103003

\bibitem[{{Sch{\"a}fer}(2009)}]{SC09.1}
{Sch{\"a}fer}, B.~M. 2009, International Journal of Modern Physics D, 18, 173

\bibitem[{{Senatore} {et~al.}(2010){Senatore}, {Smith}, \&
  {Zaldarriaga}}]{SE10.1}
{Senatore}, L., {Smith}, K.~M., \& {Zaldarriaga}, M. 2010, Journal of Cosmology
  and Astro-Particle Physics, 1, 28

\bibitem[{{Shandarin}(1980)}]{SH80.1}
{Shandarin}, S.~F. 1980, Astrophysics, 16, 439

\bibitem[{{Sugiyama}(1995)}]{SU95.1}
{Sugiyama}, N. 1995, \apjs, 100, 281

\bibitem[{{White}(1984)}]{WH84.1}
{White}, S.~D.~M. 1984, \apj, 286, 38

\bibitem[{{Xia} {et~al.}(2011){Xia}, {Baccigalupi}, {Matarrese}, {Verde}, \&
  {Viel}}]{XI11.1}
{Xia}, J.-Q., {Baccigalupi}, C., {Matarrese}, S., {Verde}, L., \& {Viel}, M.
  2011, \jcap, 8, 33

\bibitem[{{Zel'dovich}(1970)}]{ZE70.1}
{Zel'dovich}, Y.~B. 1970, \aap, 5, 84

\end{thebibliography}

\end{document}